\documentclass{pas}
\usepackage{wrapfig}
\usepackage{float}
\usepackage{multirow}
\usepackage{graphicx, subfigure}
\usepackage{subcaption}
\usepackage{multirow}
\usepackage{lineno}
\usepackage{natbib}
\usepackage[autostyle]{csquotes}
\usepackage[english]{babel}

\begin{document}

\lefttitle{Publications of the Astronomical Society of Australia}
\righttitle{Cambridge Author}

\jnlPage{1}{4}
\jnlDoiYr{2021}
\doival{10.1017/pasa.xxxx.xx}

\articletitt{Research Paper}

\title{Globular Clusters GMRT Pulsar Survey (GCGPS) III: ORION - a multi-beam acceleration search pipeline}
\author{\sn{Nishant} \gn{Deo}$^{1,2}$,
\sn{Jyotirmoy} \gn{Das}$^{1}$,
\sn{Jayanta} \gn{Roy}$^{1}$,
\sn{Saurabh} \gn{Raj}$^{1}$,
\sn{Karel} \gn{Adamek}$^{3}$,
\sn{Wesley} \gn{Armour}$^{4}$,
\sn{Bhaswati} \gn{Bhattacharyya}$^{1}$,
\sn{Jayaram} \gn{Chengalur}$^{1}$,
\sn{Paulo C. C.} \gn{Freire}$^{5}$,
\sn{Scott M.} \gn{Ransom}$^{6}$,
\sn{Sridhar} \gn{Gajendran}$^{1}$,
\sn{Sanjay} \gn{Kudale}$^{1}$,
\sn{Mekhala V.} \gn{Muley}$^{1}$,
\sn{Adityan} \gn{S}$^{5}$,
\sn{Junhao} \gn{Zhang}$^{6}$}

\affil{$^1$National Centre for Radio Astrophysics (NCRA), Pune 411007, Maharashtra, India\\
$^2$Indian Institute of Science Education and Research (IISER), Kolkata, Mohanpur 741246, West Bengal, India\\
$^3$Department of Physics, Silesian University in Opava, Opava 74601, Czech Republic\\
$^4$Oxford e-Research Centre (OeRC), University of Oxford, Oxford OX1 3PJ, United Kingdom\\
$^5$Max-Planck-Institut für Radioastronomie (MPIfR), Auf dem Hügel 69-53121, Bonn, Germany\\
$^6$National Radio Astronomy Observatory (NRAO), Charlottesville, VA, USA\\
$^7$Loyola College, Nungambakkam, Chennai 600034, Tamil Nadu, India\\
$^8$Department of Engineering Science, University of Oxford, Oxford OX1 3PJ, United Kingdom}

\citeauth{Deo N, Das J, Roy J, et al., Globular Clusters GMRT Pulsar Search (GCGPS) III: ORION - a multi-beam acceleration search pipeline. {\it Publications of the Astronomical Society of Australia} {\bf 00}, 1--12. https://doi.org/10.1017/pasa.xxxx.xx}

\history{(Received XX/XX/XX; revised XX/XX/XXXX; accepted XX/XX/XXXX)}

\begin{abstract}
Globular clusters (GCs) host diverse millisecond pulsar populations shaped by stellar encounters and binary interactions, making them valuable laboratories for studying compact-object evolution and strong-field gravity. However, the known GC pulsar population remains incomplete, motivating surveys with improved spatial coverage and efficient search strategies. We present the observing strategy for Phase~II of the Globular Clusters GMRT Pulsar Survey (GCGPS), which uses the SPOTLIGHT system to form up to 160 synthesised beams simultaneously, substantially expanding the instantaneous field of view compared with the two phased-array beams used in Phase~I. The Phase~II configuration is tunable through adjustments to the overlap between adjacent beams and the number of antennas contributing to each beam, allowing spatial coverage to be balanced against sensitivity. We quantify the sensitivity losses arising from intra-channel dispersion smearing and finite sampling to motivate the adopted dispersion-measure limits for Bands~3 and~4. To process the resulting multibeam data, we introduce \texttt{ORION}, a GPU-accelerated pulsar search pipeline combining the high-throughput, unmatched Fourier-domain filtering of \texttt{PULSCAN} with localized parameter refinement using \texttt{PRESTO}. Tests using simulated binary pulsars demonstrate processing-time reductions of approximately 15--30 times relative to the CPU-based Phase~I pipeline, while yielding comparable or higher folded detection significance for the known pulsars. On-sky validation with Cycle~49 observations recovered four known GC pulsars spanning spin periods of approximately 5.4--406~ms. Using 8 GPUs and 64 CPU cores, ORION processes full 160-beam observations in approximately 0.25–0.33 of the observation duration. Together, the Phase~II observing strategy and ORION provide a scalable framework for expanding GC pulsar searches, while future development will target improved searches for highly accelerated and jerked binary systems.
\end{abstract}

% \begin{keywords}
% Radio Astronomy, Millisecond Pulsars, Globular Clusters 
% \end{keywords}

\maketitle

\section{Introduction}

Pulsars are rapidly rotating, highly magnetised neutron stars that emit radiation powered by the loss of their rotational energy. Their rotational periods range from milliseconds to tens of seconds \citep{Backer_1982, Caleb_2022}. The serendipitous discovery of the first pulsar by \citet{Hewish_1968} marked a major milestone in both time-domain astronomy and fundamental physics. The extreme gravitational and magnetic environments of neutron stars make them unique laboratories for studying matter \citep{Chamel_2008} and gravity under conditions inaccessible on Earth, including tests of fundamental theories such as general relativity \citep{Kramer_2009, Freire_2024}. Millisecond pulsars (MSPs) are a subclass of pulsars with exceptionally short rotation periods, typically a few milliseconds, thought to be spun-up or recycled through the accretion of matter and angular momentum from a binary companion \citep{Alpar_1982, Bhattacharya_1991}. Discovering new MSPs expands the ensemble of exceptionally stable celestial clocks used in Pulsar Timing Array, thereby improving their sensitivity to nanohertz gravitational waves \citep{Hobbs_2017, Lommen_2015}.

Driven by the exceptional scientific potential of the pulsar population, there is a compelling need to design increasingly sensitive and computationally powerful pulsar surveys capable of probing previously unexplored regions of the pulsar parameter space. Early pioneering efforts, including the Parkes Multibeam Survey \citep{Manchester_2001, Lorimer_2006}, the Arecibo Pulsar Survey \citep{Cordes_2006, Hessels_2007}, and the GBT Globular Cluster Survey \citep{Ransom_2005}, established the foundations for increasingly sensitive and wide-ranging pulsar searches. More recent campaigns have further extended these capabilities, including the GBT Northern Celestial Cap survey \citep{Stovall_2014, McEwen_2024}, the GMRT High Resolution Southern Sky (GHRSS) survey  \citep{Bhattacharyya_2016, Bhattacharyya_2019, Singh_2023}, the TRAPUM survey with MeerKAT \citep{Ridolfi_2022, Turner_2024}, and the ongoing Five-hundred-meter Aperture Spherical Telescope (FAST) Galactic Plane Pulsar Snapshot (GPPS) survey \citep{Han_2021, Wang_2025}. To date, more than 4,300\footnote{\url{https://www.atnf.csiro.au/research/pulsar/psrcat/}} pulsars have been identified, including approximately 700 MSPs.

Despite these advances, the observed pulsar population represents only a fraction of the population expected in the Galaxy. Population-synthesis studies predict an active Galactic population of several hundred thousand radio pulsars \citep{Lyne_1985}, while simulations such as {\tt PsrPopPy} \citep{Bates_2014} provide a framework for modelling their detectability. A substantial population therefore remains to be discovered, particularly among pulsars that are intrinsically faint or difficult to detect because of their observational characteristics. Globular clusters (GCs) are especially compelling targets in this context, owing to their rich populations of compact objects and binary systems. The high stellar densities in these environments also make them valuable laboratories for studying pulsar populations and the dynamical processes that shape them.

The Globular Cluster GMRT Pulsar Survey (GCGPS) is specifically designed to search for pulsars in GCs using the uGMRT\footnote{\url{https://www.ncra.tifr.res.in/~jroy/GC.html}} \citep{Das_2025, Das_2026}. While GCGPS Phase I operated in Band 3 (300--500 MHz) and Band 4 (550--750 MHz) using the GMRT Wideband Backend (GWB), Phase II leverages the commensal {\tt SPOTLIGHT} system \citep{Roy_2024}. In this paper, we introduce the GCGPS Phase II using the SPOTLIGHT system and {\tt ORION}\footnote{\url{https://github.com/nishantdeo08/ORION}}, a high-performance multibeam search pipeline developed for this survey.
The remainder of this paper is organised as follows. Section~\ref{sec:science_motivation} outlines the scientific motivation for searching globular clusters for pulsars. Section~\ref{sec:gcgps_phase2} summarises the GCGPS Phase I survey and describes the Phase II observing strategy with SPOTLIGHT, including the sensitivity losses from intra-channel dispersion smearing and finite sampling that set the adopted dispersion-measure limits, and the advantages of multibeam searches. Section~\ref{sec:orion} presents the \texttt{ORION} pipeline and its four stages: candidate generation, candidate reduction, parameter refinement, and candidate folding and classification. Section~\ref{sec:orion_validation} validates \texttt{ORION} by comparing its processing time and detection significance with the Phase I pipeline using simulated binary pulsars, quantifying its computational cost, and demonstrating the on-sky recovery of known globular-cluster pulsars in Cycle 49 observations. Section~\ref{sec:scientific_implications} discusses the scientific implications for dynamically processed binaries, Section~\ref{sec:future} describes planned developments, in particular \texttt{\textit{Jerk}$\_$PULSCAN}, and Section~\ref{sec:summary} summarises our conclusions.

\section{Science Motivation}
\label{sec:science_motivation}

Globular clusters (GCs) are old ($\sim10$--$13$~Gyr), dense,
gravitationally bound stellar systems in which stellar encounters occur
frequently enough to shape the formation and evolution of binary
systems. Because the radio-loud lifetime of an ordinary pulsar is far
shorter than the typical age of the observed GCs, pulsars cannot constitute a significant primordial population; the MSPs observed today instead form through binary recycling, in which a neutron star spins up by accreting matter and angular momentum from a low-mass companion in a low-mass X-ray binary \citep{Alpar_1982, Bhattacharya_1991}. The high stellar
densities in GC cores make binary formation, exchange, hardening, and
disruption far more efficient than in the Galactic field, producing a
much larger MSP population per unit stellar mass than is found among
field pulsars \citep{Freire_2013}. Unlike field MSPs,
whose orbital properties largely reflect binary evolution alone, GC
binaries can be subsequently reshaped by exchange encounters and other
dynamical interactions, so their orbital parameters encode a record of
the cluster's dynamical history \citep{Freire_2013, Verbunt_2014}. Moreover, the
same dynamical environment can reshape these binaries
through repeated stellar encounters, producing orbital configurations
and companion properties that are uncommon among field MSPs. Thus, GC
MSPs provide not only a larger population of recycled pulsars, but also
a distinct laboratory for studying binary evolution under strong
dynamical processing.

\subsection{Diverse and Dynamically Processed MSP Populations}
\label{sec:gc_pulsar_populations}

GC MSPs span a wide range of binary configurations: low-eccentricity
MSP--white-dwarf binaries produced by largely undisturbed recycling;
compact ``spider'' systems (black widows and redbacks), in which
energetic emission from the pulsar ablates its companion
\citep{Roberts_2013}; and eccentric, dynamically assembled systems that
are rare in the Galactic field. Exchange encounters can replace an
MSP's original companion with a more massive one or substantially alter
its orbit \citep{Sigurdsson_1993}, while repeated interactions can raise
the eccentricity of an existing binary or disrupt it entirely,
producing isolated MSPs \citep{Verbunt_2014, Ridolfi_2021}.

The observed population illustrates how extreme these effects can be.
PSR~J0514$-$4002A (NGC~1851A) is a $\sim5$-ms pulsar in an eccentric
($e\simeq0.89$), 18.8-d orbit with a massive companion, enabling
relativistic mass measurements difficult to reconcile with isolated
binary evolution \citep{Freire_2008, Ridolfi_2019}. NGC~1851E hosts a
$\sim5.6$-ms pulsar in a $\sim7$-d eccentric binary with total mass
$\sim3.9\,M_\odot$, placing its companion in the poorly understood neutron-star--black-hole mass gap \citep{Barr_2024}. GCs have also
produced the most extreme known neutron-star spin rate, PSR~J1748$-$2446ad
in Terzan~5, at $\simeq716$~Hz \citep{Hessels_2006}. The first
GCGPS discovery, J1617$-$2258A in NGC~6093 -- an MSP in a compact,
eccentric ($e\simeq0.54$) 19-h binary \citep{Das_2025} -- shows that
unusual configurations continue to emerge even after decades of
targeted GC pulsar searches, motivating continued, sensitive surveys.

\subsection{Dynamical Environment and Spatial Distribution}
\label{sec:gc_dynamics}

The diversity of GC MSPs is closely tied to the dynamical environment of their host clusters. Unlike field MSPs, whose orbital properties are primarily determined by binary evolution and recycling, GC binaries can undergo repeated encounters with other stars. These interactions can exchange companions, modify orbital eccentricities, harden or disrupt binaries, and produce isolated MSPs or compact spider systems. As a result, the GC environment can generate MSP configurations that are rare or absent among the normal Galactic field population.

The extent of this processing depends on the cluster's encounter rate. The total stellar encounter rate, $\Gamma$, and the encounter rate per unit mass, $\gamma = \Gamma/M_{\rm tot}$, are commonly used to characterise, respectively, how readily a cluster forms interacting binaries and how likely an existing binary is to be further reprocessed \citep{Verbunt_1987, Pooley_2003, Freire_2011}. Clusters with high encounter rates, such as Terzan~5, M28 (NGC~6626), and NGC~6440, provide environments in which repeated stellar encounters can substantially modify MSP binaries, producing eccentric systems, exchange products, and isolated MSPs \citep{Freire_2013, Ridolfi_2021}. For comparison, the MSP population of the lower-encounter-rate cluster 47~Tuc is dominated by relatively low-eccentricity binaries, although dynamical interactions are still important there \citep{Freire_2013}. Comparing populations across clusters with different $\Gamma$ and $\gamma$ therefore tests how binary evolution responds to the dynamical environment.

% Clusters with high
% $\Gamma$ and $\gamma$ are expected to host a larger fraction of
% dynamically modified binaries -- eccentric systems, exchange products,
% and spiders -- while lower-encounter-rate clusters preserve binaries
% closer to the products of standard recycling \citep{Freire_2013,
% Ridolfi_2021}. 

Because encounter rates peak in the cluster core, pulsars nearer the
centre are more likely to be dynamically processed, while pulsars at
larger radii -- out to and beyond the half-light radius, $r_{\rm h}$ --
can include less-disturbed binaries and spider systems that received a
recoil kick from the encounters that formed them \citep{Freire_2011,
Ridolfi_2021}. MSPs are therefore not necessarily confined to
the core, and restricting a survey to the central region can miss
pulsars at larger cluster-centric radii. Surveys covering each cluster
out to at least the half-light radius, $r_{\rm h}$, are consequently
important for building a more complete MSP census and for studying the
spatial distribution of the population. The pulsar position is also
essential for interpreting timing measurements, since the line-of-sight
acceleration depends on the pulsar's location within the cluster
potential.

% Mass segregation and interaction kicks mean MSPs are not
% confined to the core, so surveys covering each cluster out to at least
% $r_{\rm h}$ are needed for a representative, spatially complete sample
% -- important both for population statistics and because the
% line-of-sight accelerations inferred from timing depend on the
% pulsar's location within the cluster potential.

\subsection{The Incomplete Globular Cluster MSP Population}

Despite decades of targeted searches, the known GC MSP population remains far
from complete. Realistic $N$-body models of cluster evolution predict that a
typical GC can host $\sim$10--20 MSPs, and the most massive clusters close to
a hundred \citep{Ye_2019}, while unresolved $\gamma$-ray emission from several GCs independently points to tens of undetected MSPs per cluster
\citep{Abdo_2010}. Against these predictions, the observed population --
currently several hundred MSPs across fewer than 50 clusters -- represents
only a fraction of what is thought to exist, limited primarily by survey
sensitivity and by the many GCs that have never been searched with
instruments sensitive enough to detect their fainter, more distant, or
steeper-spectrum members. This gap between predicted and detected
populations is itself a central motivation for continued, sensitive surveys:
each undiscovered MSP is a potential outlier at the extremes of spin,
eccentricity, or companion mass -- precisely the systems most useful for the
applications described above.

\subsection{Science Enabled by GC Pulsars}
\label{sec:gc_pulsar_applications}

A larger, well-timed GC pulsar population enables science beyond the
census itself. Relativistic binaries, especially the eccentric,
massive systems favoured by dynamical formation,  allow precision
measurement of post-Keplerian parameters and stringent tests of gravity
in the strong-field regime, as demonstrated by NGC~1851A
\citep{Dutta_2025, Ridolfi_2019, Freire_2011}. Extreme individual systems constrain
neutron-star physics directly: the 716-Hz spin of the Terzan~5 pulsar constrains neutron-star structure/equation-of-state models \citep{Hessels_2006}, and the mass-gap companion in NGC~1851E offers a rare probe of compact-object
formation \citep{Barr_2024}. Timing-derived accelerations across a
spatially complete pulsar sample can further constrain a cluster's
central mass density, mass-to-light ratio, and the distribution of dark
remnants \citep{Prager_2017, Perera_2017}, while multiple pulsar
sightlines through the same cluster provide DM and rotation-measure
constraints on its ionised gas and magnetic field \citep{Freire_2013,
Abbate_2020}.

The most valuable discoveries are therefore not necessarily the most
numerous, but the systems at the extremes of spin, eccentricity,
companion mass, and cluster location, extremes that are set by, and
trace, the dynamical environment of the host cluster. This motivates
sensitive, spatially complete pulsar surveys of GCs, particularly
clusters combining high encounter rates, large angular extent, and
incomplete existing censuses: the goal of the GCGPS survey and the
ORION search pipeline described in this paper.

\section{Globular Cluster GMRT Pulsar Survey}
\label{sec:gcgps_phase2}

In this section, we give a brief overview of the Phase I of the GCGPS survey. We then motivate GCGPS Phase II and provide a brief comparison with Phase I.  

\subsection{Survey Overview and Phase I}
\label{sec:gcgps_phase1}

\begin{figure*}
    \centering
    \includegraphics[width=\textwidth]{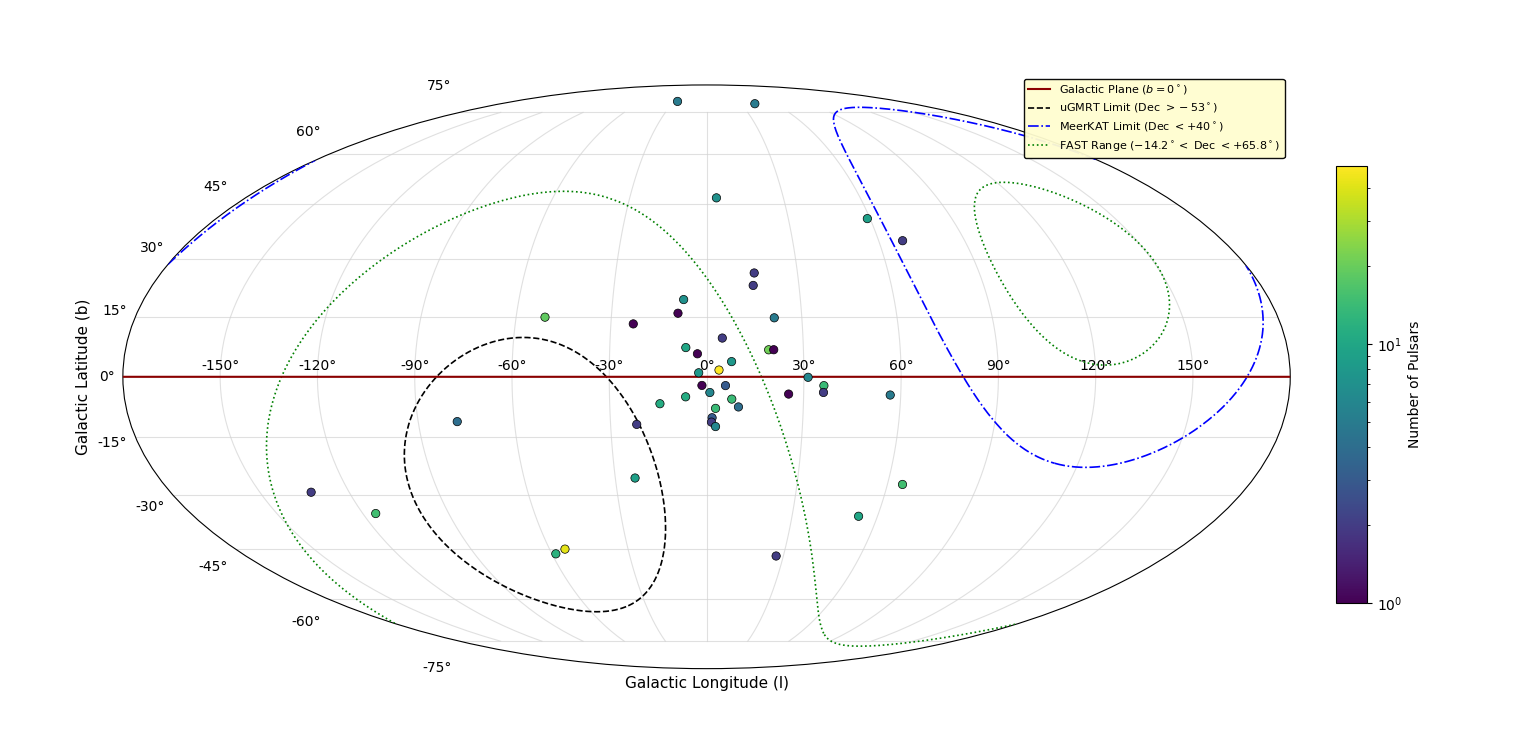}
    \caption{The Galactic distribution of the 46 globular clusters in
    which pulsars, predominantly millisecond pulsars (MSPs), have been
    detected. Of these, 41 GCs are accessible to the uGMRT.}
    \label{fig:pulsar_galactic_gc}
\end{figure*}

The Globular Clusters GMRT Pulsar Survey (GCGPS) is a dedicated survey
aimed at discovering pulsars in Galactic globular
clusters (GCs) using the upgraded Giant Metrewave Radio Telescope
(uGMRT) \citep{Das_2025, Das_2026}. The survey targets GCs accessible
to the uGMRT that have not previously been searched with comparable
sensitivity, primarily using Band-3 (300--500\,MHz) and Band-4
(550--750\,MHz) with the GMRT Wideband Backend (GWB). The low-frequency
coverage of the uGMRT provides a complementary approach to existing GC
pulsar surveys conducted primarily at higher radio frequencies
\citep{Das_2025}. Figure~\ref{fig:pulsar_galactic_gc} shows the the Galactic distribution of the GCs in which pulsars have been detected.

The observations employed two simultaneously formed
phased-array (PA) beams, using the central-square and third-arm
configurations to provide complementary sensitivity and spatial
coverage. Interferometric data were recorded simultaneously with the
beamformed observations, enabling subsequent localization of pulsar
candidates. Phase~I has so far resulted in the discovery of seven MSPs
in four GCs where no pulsars were previously known
\citep{Das_2025, Das_2026}.

\subsection{SPOTLIGHT and GCGPS Phase II}
\label{sec:spotlight_phase2}

GCGPS Phase~II extends the Phase~I observing strategy by exploiting the multibeam capability of the SPOTLIGHT backend \citep{Roy_2024}. SPOTLIGHT enables the formation of a large number of synthesized post-correlation (PC) beams \citep{Roy_2015}. For
GCGPS Phase~II, we plan to use 160 PC beams to provide substantially larger instantaneous spatial coverage than the two phased-array beams used in Phase~I \citep{Das_2026}. Unlike the fixed beam configuration
adopted in Phase~I, the Phase~II observing strategy is designed to be tunable to the angular extent of the target globular cluster. In particular, both the overlap between adjacent beams and the number of antennas contributing to each phased beam can be adjusted according to the required balance between field of view and sensitivity.
The following configuration will be for the survey:
\begin{itemize}
    \item Sampling time 327.68 $\mu$s
    \item 1k Frequency Channels
\end{itemize}
 
The coarser time and frequency resolution relative to Phase~I reduces the sensitivity to short-period and highly dispersed pulsars. We therefore quantify the theoretical sensitivity loss from intra-channel dispersion smearing and finite sampling for this configuration (Figure~\ref{fig:snr_retention}), and use it to define the DM range of the survey.

\subsubsection{Dispersion and intra-channel smearing}
\label{sec:dispersion_smearing}

Even after dedispersion at the correct trial DM, the finite width of each frequency channel leaves a residual intra-channel dispersion smear,
\begin{equation}
t_{\rm DM} \simeq 8.3~\mu{\rm s}
\left(\frac{DM}{\mathrm{pc\,cm^{-3}}}\right)
\left(\frac{\Delta\nu}{\mathrm{MHz}}\right)
\left(\frac{\nu}{\mathrm{GHz}}\right)^{-3},
\label{eq:intra_channel_smearing}
\end{equation}
where $\Delta\nu$ is the channel width and $\nu$ is the observing frequency \citep{CordesMcLaughlin2003}. For the Phase~II configuration, 1024 channels span 200~MHz, giving $\Delta\nu \simeq 0.195$~MHz in both Band~3 (300--500~MHz) and Band~4 (550--750~MHz).

This smearing adds in quadrature with the other broadening terms,
\begin{equation}
W_{\rm obs} =
\left(
W_{\rm int}^2 +
t_{\rm samp}^2 +
t_{\rm DM}^2 +
t_{\Delta DM}^2 +
t_{\rm scat}^2
\right)^{1/2},
\label{eq:effective_pulse_width}
\end{equation}
and, for a periodic search, degrades the recoverable S/N through the standard radiometer scaling \citep{LorimerKramer2005}
\begin{equation}
\mathrm{S/N} \propto
\left(\frac{P-W_{\rm obs}}{W_{\rm obs}}\right)^{1/2}.
\label{eq:periodic_snr}
\end{equation}
Isolating the intra-channel term ($W_{\rm obs} = (W_{\rm int}^2+t_{\rm DM}^2)^{1/2}$) and normalising to the smearing-free case gives the fractional S/N retained,
\begin{equation}
\frac{(\mathrm{S/N})}{(\mathrm{S/N})_0}
=
\left[
\frac{(P-W_{\rm obs})\,W_{\rm int}}
{(P-W_{\rm int})\,W_{\rm obs}}
\right]^{1/2}.
\label{eq:snr_retention}
\end{equation}

\begin{figure*}
\includegraphics[width=0.50\textwidth]{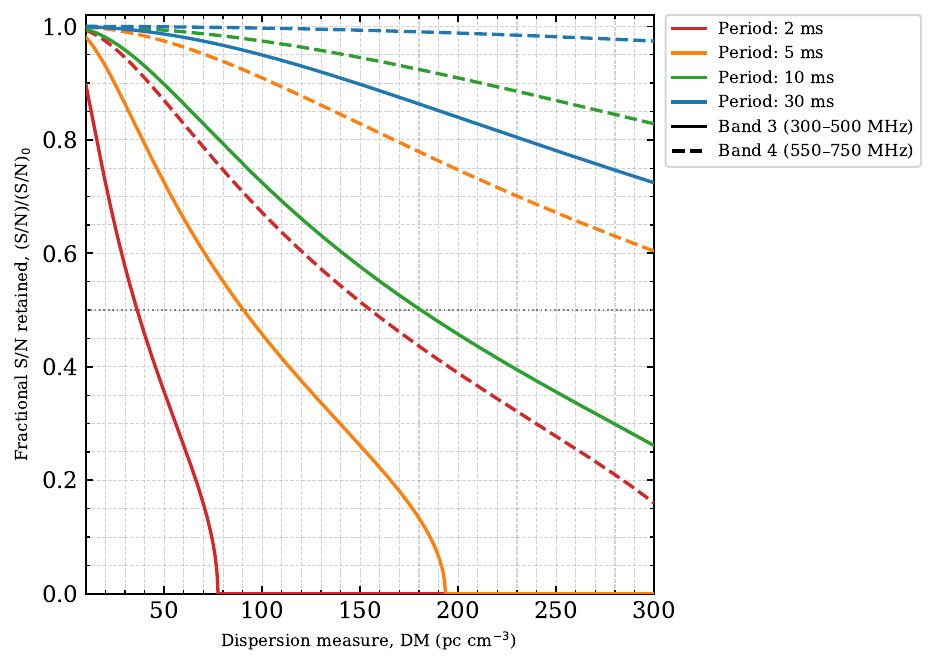}
\includegraphics[width=0.47\textwidth]{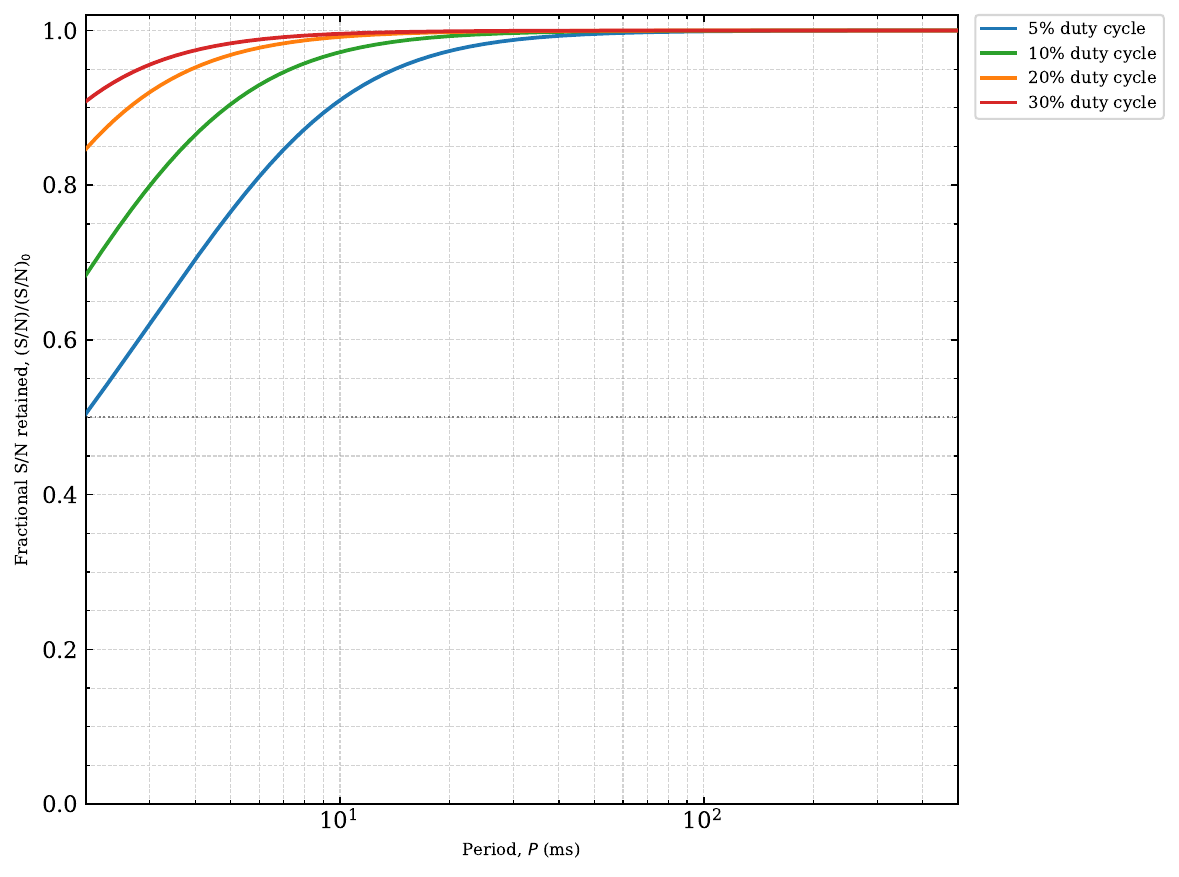}
\caption{Fractional search S/N retained with 327.68 $\mu$s and 1k channels. a) The sensitivity loss occurring due to intra-channel dispersion for MSPs in Band 3, Band 4. b) The sensitivity loss due to the sampling time for different duty cycle pulsars.}
\label{fig:snr_retention}
\end{figure*}

We evaluate Equation~\ref{eq:snr_retention} at the central frequency of each band (400~MHz for Band~3, 650~MHz for Band~4) for pulsars with periods of 2, 5, 10, and 30~ms, adopting a duty cycle of 20\%, representative of the range typically observed for recycled and millisecond pulsars \citep[$\sim$10--30\%, in contrast to the $\sim$5\% typical of the normal pulsar population;][]{Kramer1998}. Figure~\ref{fig:snr_retention} shows the resulting S/N retention as a function of DM. Band~4 retains substantially more sensitivity than Band~3 at fixed DM, since $t_{DM}\propto\nu^{-3}$: sub-10~ms pulsars lose more than half their periodic S/N by DM~$\sim$~100--200~pc\,cm$^{-3}$ in Band~3, whereas the 30 ms pulsar remains comparatively unaffected across the full DM range considered.

The finite sampling interval also broadens the observed pulse profile and consequently reduces the sensitivity to narrow pulses. For the Phase~II configuration, the sampling interval is 327.68~$\mu$s. To isolate this effect from dispersion smearing, we calculate the effective pulse width as
\begin{equation}
W_{\rm obs} =
\left(W_{\rm int}^2+t_{\rm samp}^2\right)^{1/2},
\end{equation}
with $W_{\rm int}=0.2P$. The corresponding fractional S/N retention is then calculated using Equation~\ref{eq:snr_retention}.

Because the intrinsic width itself is uncertain, we also test how sensitive this result is to the assumed duty cycle. Figure~\ref{fig:snr_retention} shows the S/N retention for the same set of periods, now evaluated at a fixed representative frequency (400~MHz) while the duty cycle is varied from 5\% to 30\%, spanning the range reported for millisecond and mildly recycled pulsars \citep{Kramer1998}. As expected, a smaller assumed duty cycle steepens the S/N loss with period, since $t_{DM}$ then represents a proportionally larger fraction of $W_{obs}$; this effect is most pronounced for the shortest-period pulsars, for which $W_{int}$ is already comparable to or smaller than $t_{DM}$ over much of the DM range explored. However, since MSPs tend to have higher duty cycles, the S/N loss is not significant as seen in Figure~\ref{fig:snr_retention}.

\iffalse
\begin{figure}
\centering
\includegraphics[width=\columnwidth]{Plots/gcgps_phase2_dispersion_snr_vs_period_duty.pdf}
\caption{Fractional search S/N retained with the 327.68$\mu$s as a function of period for varying duty-cycles (pulse-widths)}
\label{fig:sampling_snr}
\end{figure}

Figure~\ref{fig:sampling_snr} shows the resulting S/N retention as a function of pulsar period. The sampling interval has a substantial effect on the detectability of millisecond pulsars, while its contribution becomes negligible for pulsars with periods much longer than the sampling interval.
\fi

Based on this analysis, we limit the Phase~II search to $\mathrm{DM}\leq50$~pc\,cm$^{-3}$ for Band 3 and $\mathrm{DM}\leq150$~pc\,cm$^{-3}$ for Band 4. Within this range, Band~3 retains $\gtrsim$90\% of the S/N for $P\gtrsim10$~ms (and $\approx$98\% for $P\gtrsim30$~ms), and Band~4 retains $\gtrsim$80\% for $P\geq5$~ms. All GCs proposed for GCGPS Phase~II lie within this DM range.

\subsection{Advantages of Multibeam Searches}
\label{sec:multibeam_advantages}

\iffalse
The Phase~II configuration provides several advantages over the single- or dual-beam observations employed in Phase~I. These include a substantially larger instantaneous field of view, flexible beam geometry, a larger number of coherently combined antennas, and the planned use of in-field phasing. Together, these capabilities allow the observing configuration to be adapted to the angular extent and pulsar distribution of individual globular clusters.
\fi
Compared with the single- or dual-beam observations of Phase~I, the Phase~II configuration provides a substantially larger instantaneous field of view and a flexible beam layout, allowing the observing setup
to be tailored to the angular extent and pulsar distribution of each
globular cluster.

\iffalse
\begin{figure*}
\centering
\includegraphics[width=\textwidth]{Plots/multibeam_map_0.5.pdf}
\caption{Multibeam tiling for M28 (NGC~6626). The 160 Phase~II PC beams are formed using 30 antennas and are shown together with the Phase~I central-square and third-arm beam responses and the positions of the known pulsars. The Phase~I configurations use 14 and 22 antennas, respectively.}
\label{fig:NGC6626_multibeam_tiling}
\end{figure*}
\fi

\begin{figure*}
\centering
\includegraphics[width=0.8\textwidth]{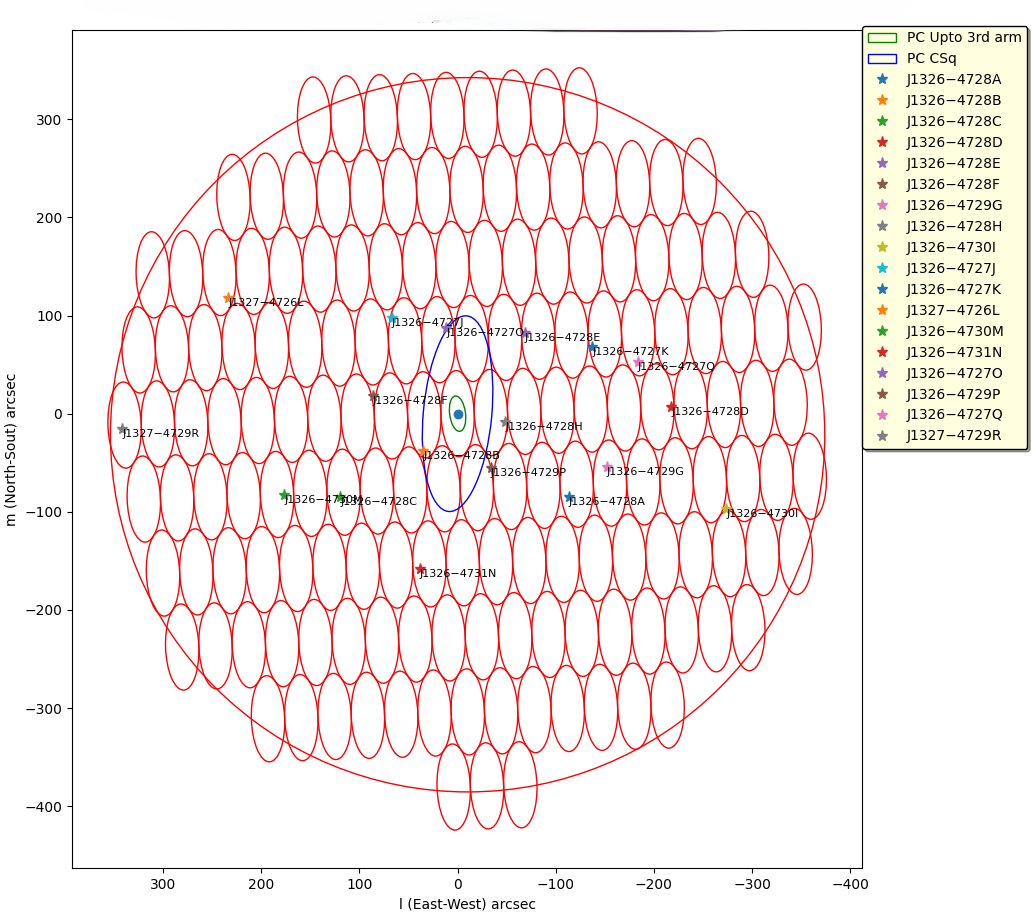}
\caption{Multibeam tiling for $\omega$~Centauri using 17 antennas per PC beam. The positions of the known pulsars are shown. The reduced array size produces broader beams, allowing the 160-beam configuration to cover the large angular extent of the cluster.}
\label{fig:Omega_Centauri_multibeam_tiling}
\end{figure*}

\subsubsection{Enhanced spatial coverage}
\label{sec:enhanced_spatial_coverage}

The increased field of view provided by SPOTLIGHT is particularly valuable for pulsar searches in globular clusters, whose angular extent can be comparable to or larger than the useful field of view of a single phased-array beam. A single pointing can therefore provide non-uniform sensitivity across a cluster, requiring multiple pointings to obtain adequate spatial coverage.

Phase~II addresses this limitation by simultaneously forming up to 160 coherently beamformed beams across the target field. This provides much finer spatial sampling than the limited number of beam responses available in Phase~I and increases the probability that a pulsar is located close to the centre of at least one synthesized beam. The relative responses of neighbouring beams can also provide useful positional information prior to interferometric localisation.

\begin{figure}
\centering
\includegraphics[width=\columnwidth]{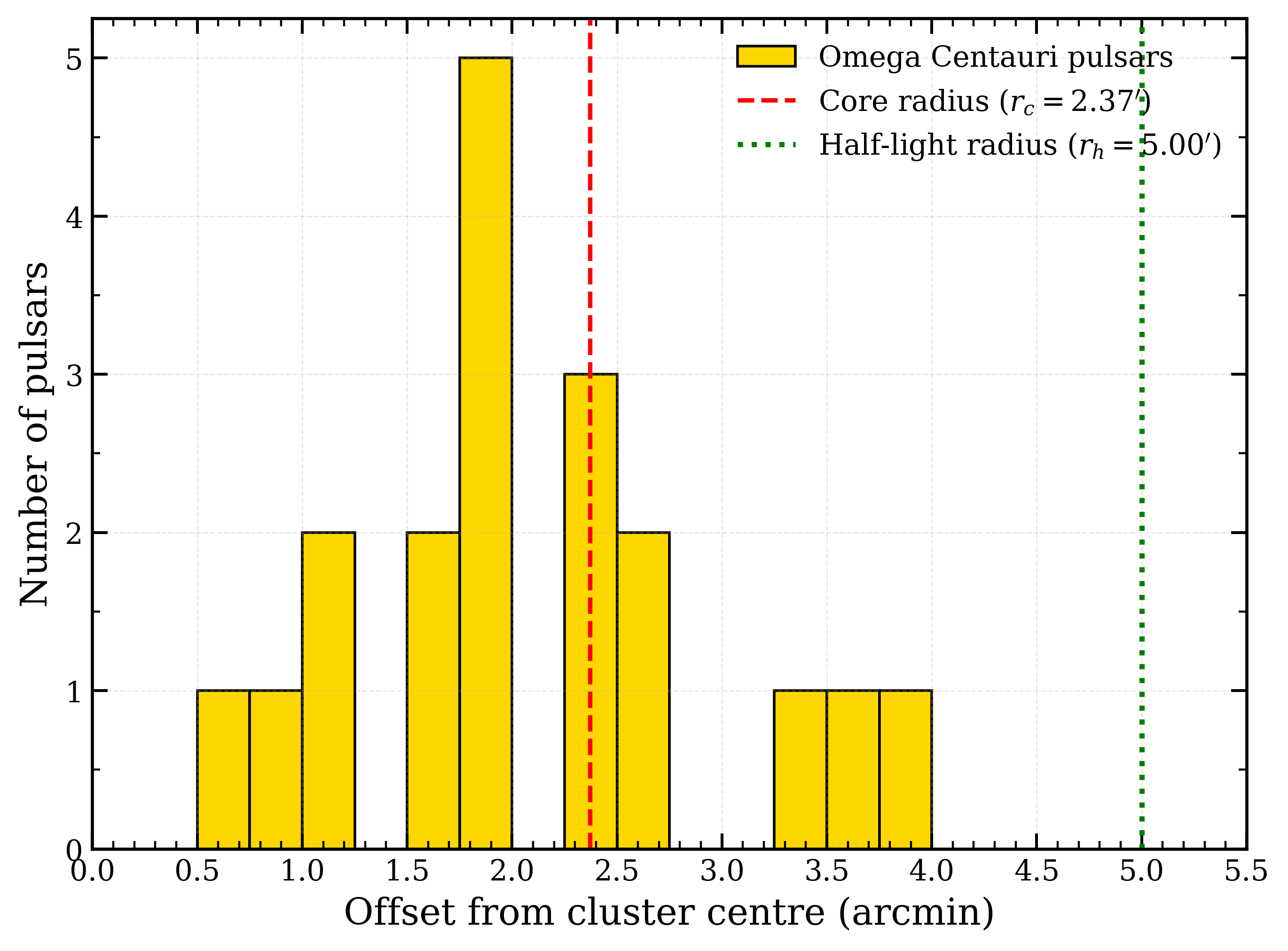}
\caption{The histogram shows the offset of pulsars in Omega Centauri. The red vertical line indicates the core-radius and the green vertical line indicates the half-light radius of the globular cluster.}
\label{fig:Omega_centauri_pulsar_radial_distribution}
\end{figure}

Figure~\ref{fig:Omega_Centauri_multibeam_tiling} shows the Phase~II multibeam configuration for $\omega$~Centauri. The known pulsars span $\sim0.5'$--$4'$ from the cluster centre, with several lying beyond the core radius ($r_c=2.37'$) and extending toward the half-light radius ($r_h=5.00'$), as shown in Figure~\ref{fig:Omega_centauri_pulsar_radial_distribution}. This demonstrates the need for sensitivity beyond the central region when searching for a complete pulsar population.

For $\omega$~Centauri, 17 antennas are used per PC beam, producing broader beams that allow the 160-beam pattern to cover a larger instantaneous area. This illustrates the flexibility of the Phase~II configuration, where the number of antennas per beam can be adjusted to balance spatial coverage and coherent sensitivity according to the angular extent of the target cluster.

Similar multibeam strategies have been employed in large-scale GC surveys with MeerKAT, including the TRAPUM survey, demonstrating the value of simultaneous coherent beams for efficiently searching spatially extended clusters \citep{Ridolfi_2022, Abbate_2022, Chen_2021}. The SPOTLIGHT implementation provides a complementary approach at the lower observing frequencies of the uGMRT.

\section{ORION: A GPU-Accelerated Multibeam Pulsar Search Pipeline}
\label{sec:orion}

The SPOTLIGHT backend can produce up to $\sim160$ simultaneously formed synthesised beams for a targeted observation. Processing these beams independently with a conventional CPU-based pulsar-search pipeline would generate a large number of intermediate candidates and incur a substantial computational cost. To address this, we developed \texttt{ORION}, a GPU-accelerated pulsar-search pipeline designed specifically for the multibeam observations of GCGPS Phase~II.

\begin{figure}
    \centering
    \includegraphics[width=0.8\linewidth]{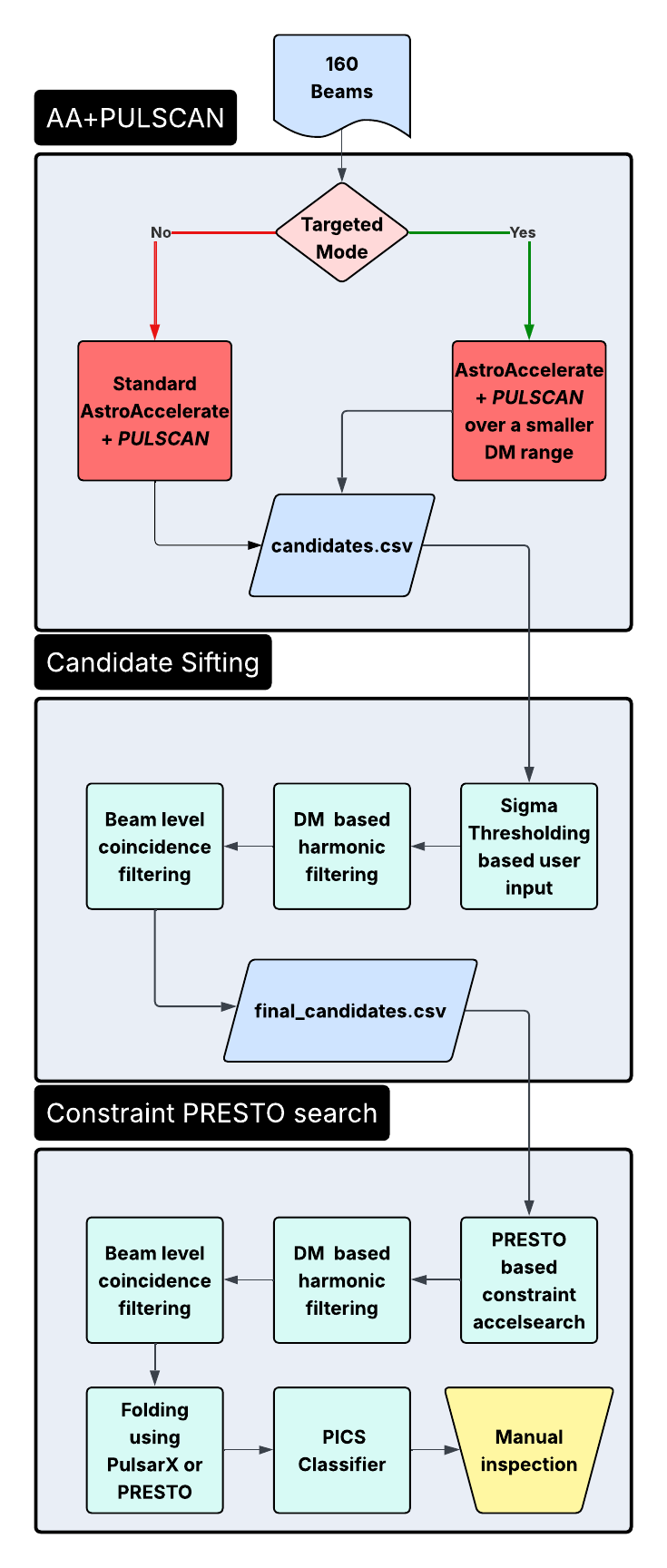}
    \caption{Overview of the \texttt{ORION} multibeam pulsar-search pipeline. The 160 beams are processed in parallel using GPU-accelerated \texttt{PULSCAN}, followed by candidate thresholding, harmonic filtering, and multibeam coincidence filtering. Surviving candidates are refined using frequency localized \texttt{PRESTO} searches. Candidates are subsequently folded, automatically classified, and manually inspected.}
    \label{fig:ORION_Flowchart}
\end{figure}

The pipeline is structured into four main stages: (i) candidate generation, (ii) candidate reduction, (iii) parameter refinement, and (iv) candidate folding and classification. An overview is shown in Figure~\ref{fig:ORION_Flowchart}. The initial search is performed using the GPU implementation of \texttt{PULSCAN} \citep{White_2025}, with the SPOTLIGHT PC-beams processed in parallel. The resulting candidates are reduced using significance, harmonic, and multibeam coincidence filtering before a frequency localized \texttt{PRESTO} search is performed around the parameters returned by \texttt{PULSCAN}. The refined candidates are subsequently folded and ranked using automated classifiers before manual inspection.

\subsection{Candidate Generation}
\label{sec:orion_candidate_generation}

The first stage of \texttt{ORION} performs a GPU-accelerated Fourier-domain acceleration search using \texttt{PULSCAN} \citep{White_2025}. Conventional Fourier-domain acceleration searches (FDAS) use matched filtering to recover the frequency drift produced by the line-of-sight acceleration of a binary pulsar. While this provides optimal sensitivity for the assumed signal model, the computational cost increases rapidly when a large acceleration range must be searched over many beams. \texttt{PULSCAN} instead uses an unmatched boxcar convolution in the Fourier domain, providing a high-throughput first-pass search \citep{White_2025}.

In \texttt{ORION}, the search is performed independently for each beam and produces candidate estimates of spin frequency, dispersion measure (DM), acceleration, and detection significance. The initial search is intentionally broad, resulting in a large candidate set that is subsequently reduced before the more computationally expensive matched-filtering stages.

For targeted searches, the known or expected DM range can be used to restrict the initial search. This reduces the parameter space explored by \texttt{PULSCAN} while retaining the same processing framework. The resulting candidate lists form the input to the subsequent reduction stages.

\subsection{Candidate Reduction}
\label{sec:orion_candidate_reduction}

The initial \texttt{PULSCAN} output contains multiple detections of the same underlying signal arising from different trial parameters, harmonic responses, and, for multibeam observations, neighbouring synthesised beams. \texttt{ORION} therefore applies a sequence of candidate-reduction steps based on detection significance, harmonic relationships, and spatial coincidence.

\subsubsection{Threshold and harmonic filtering}
\label{sec:orion_harmonic}

Candidates below a user-defined detection significance are first removed. For the analysis presented here, we adopt a threshold of $5\sigma$, with candidate periods restricted to $1$--$1000\,\mathrm{ms}$. These limits are configurable according to the time resolution and scientific requirements of a given observation.

A periodic signal can produce detections at integer multiples or submultiples of its fundamental frequency. \texttt{ORION} therefore groups candidates related by harmonic structure and retains the highest-significance representative. This is performed both among harmonics within the same DM trial and across neighbouring DM trials, substantially reducing duplicate detections while preserving the strongest representation of each periodic signal. These operations are performed independently for each beam and in parallel.

\begin{figure*}
    \centering

    \subfigure[Stage 0: Initial candidates from the raw
    \texttt{AstroAccelerate PULSCAN} output.]{
        \includegraphics[width=0.48\textwidth]{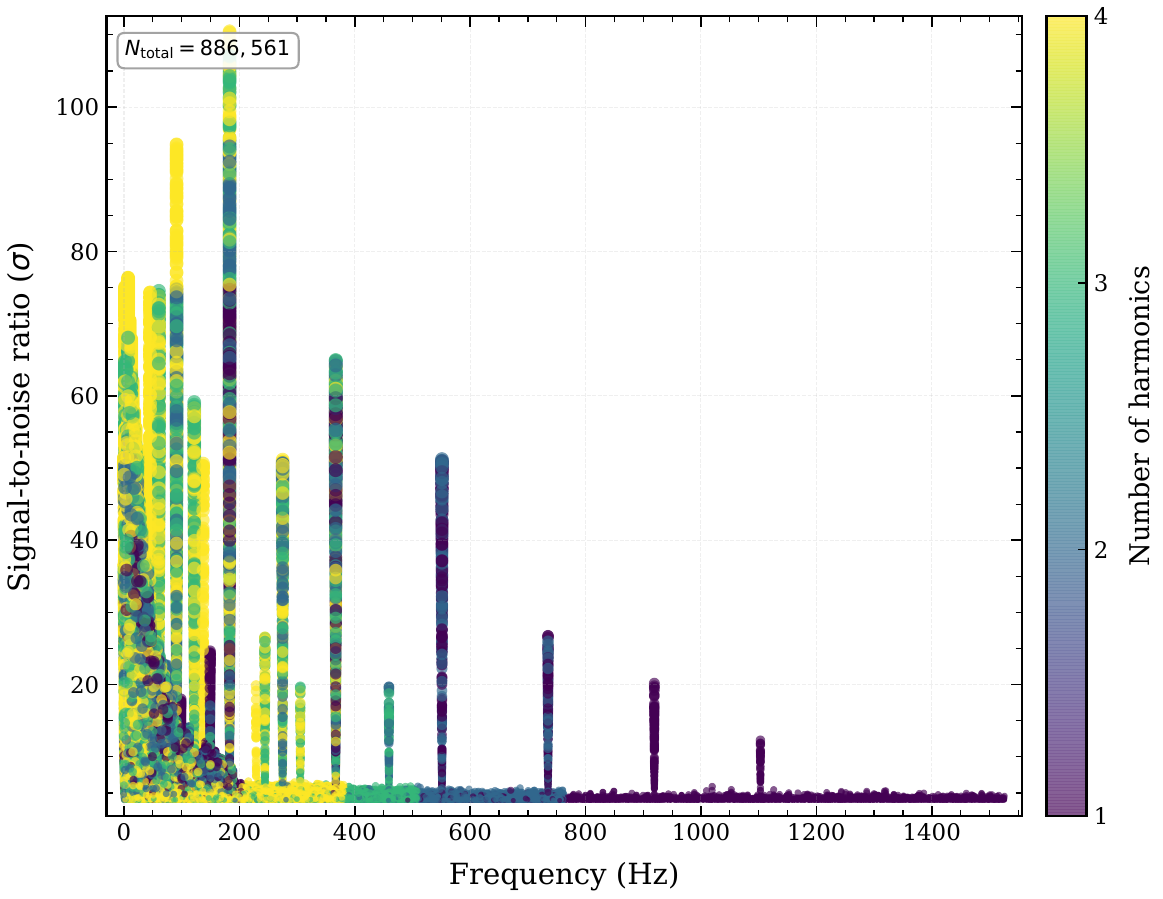}
    }
    \hfill
    \subfigure[Stage 1: Candidates after harmonic filtering.]{
        \includegraphics[width=0.48\textwidth]{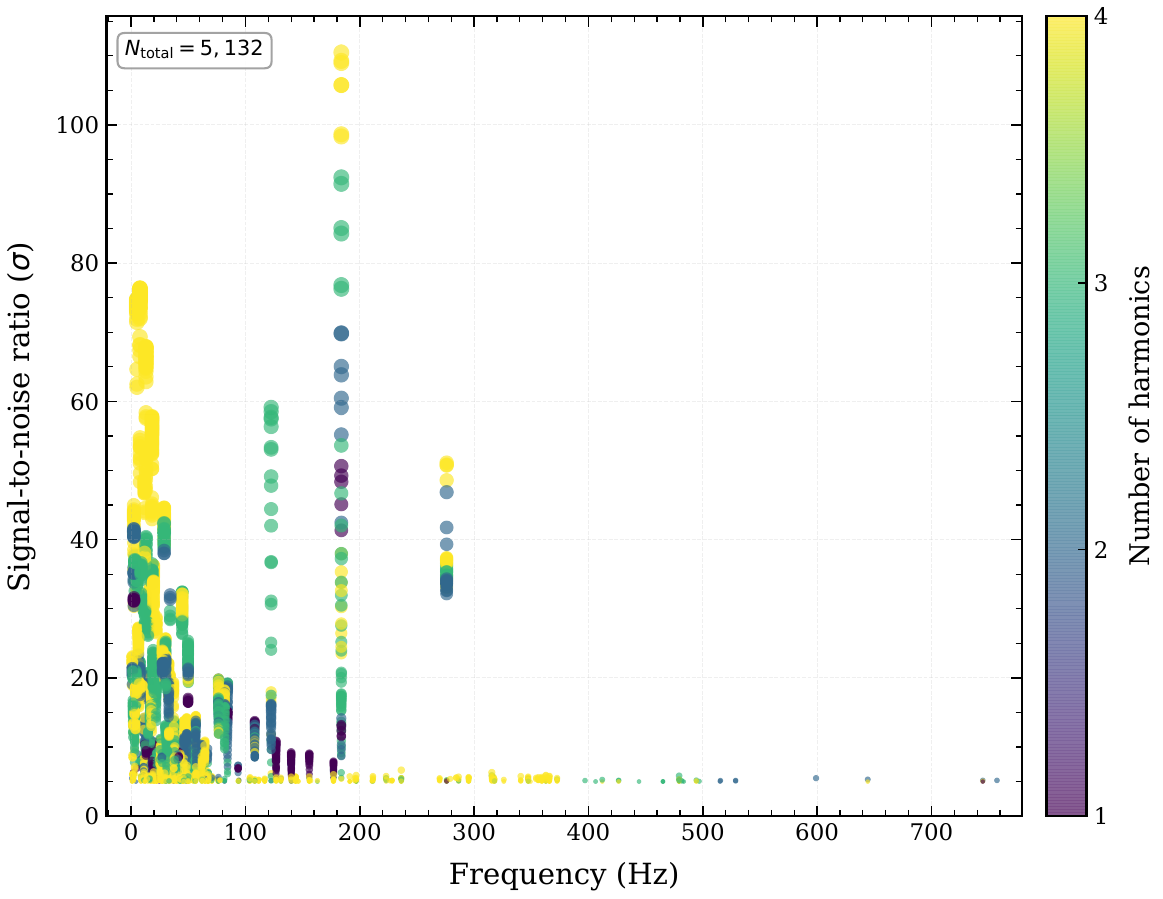}
    }

    \vspace{0.5em}

    \subfigure[Stage 2: Candidates after beam-level coincidence
    filtering.]{
        \includegraphics[width=0.48\textwidth]{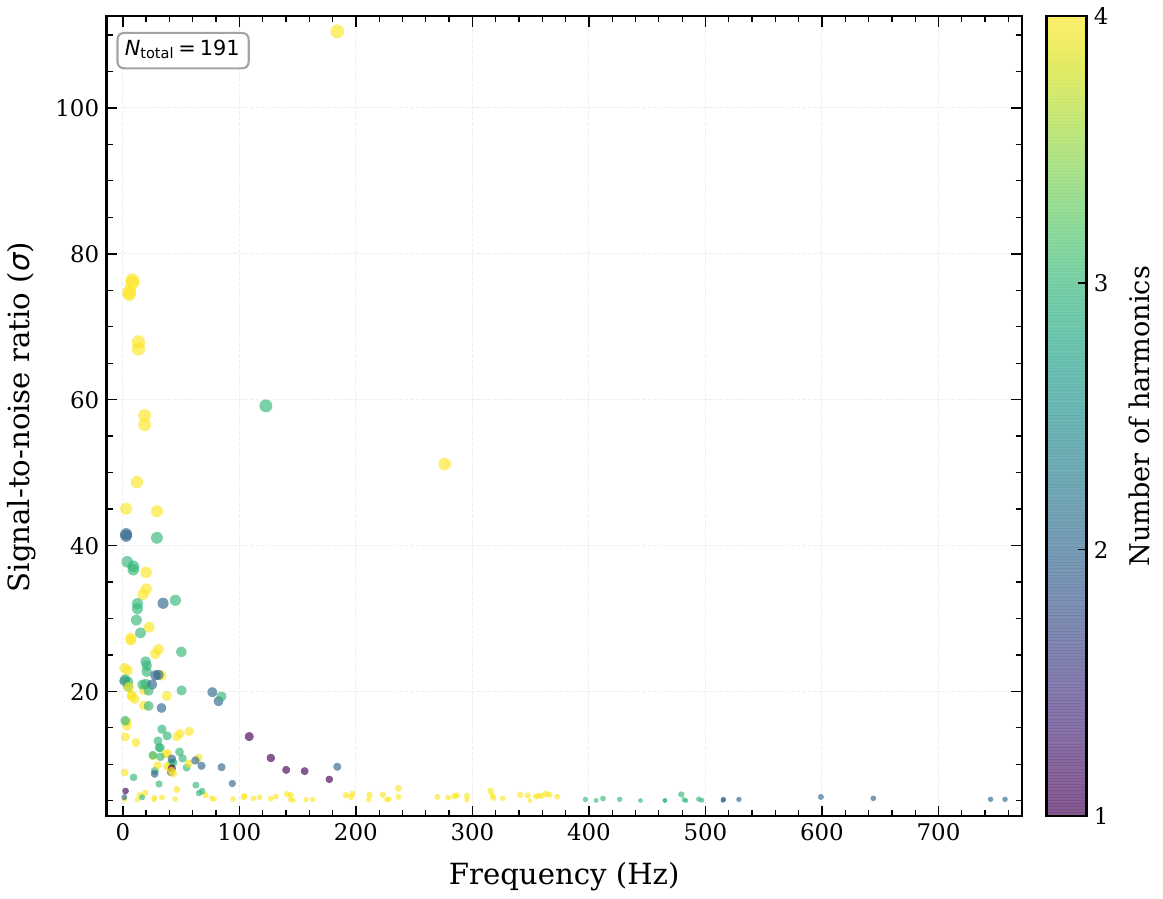}
    }

    \caption{Candidate distributions at successive stages of the \texttt{ORION} candidate-selection process for beam~0. The initial \texttt{PULSCAN} output contains a large number of detections and harmonic responses. Harmonic filtering removes redundant periodicity detections, while the subsequent beam-level coincidence filtering uses the spatial information provided by the SPOTLIGHT multibeam observation to identify consistent candidates.}
    \label{fig:ORION_candidates}
\end{figure*}

Figure~\ref{fig:ORION_candidates} illustrates the evolution of the candidate distribution through these processing stages for one beam. The initial \texttt{PULSCAN} output contains a large population of detections distributed across frequency and significance, with multiple harmonic responses visible as repeated structures. Following harmonic filtering, the candidate population is substantially condensed around the remaining independent periodicity detections. The subsequent beam-level filtering further selects candidates that are spatially consistent with the multibeam response.

\subsubsection{Multibeam coincidence filtering}
\label{sec:orion_coincidence}

The final candidate-reduction step exploits the spatial information provided by the beam pattern. A celestial source can appear in several neighbouring beams, with the detection significance varying according to its position relative to the individual beam responses. \texttt{ORION} therefore groups candidates that are consistent in frequency, DM, and other search parameters and occur in spatially neighbouring beams.

When the same signal is detected in multiple neighbouring beams, the highest-significance detection is retained as the representative candidate. Candidates detected in only a single beam are retained. Signals appearing simultaneously in multiple unrelated beams are flagged separately, since such behaviour can be indicative of instrumental or radio-frequency-interference features rather than a localized celestial source. The resulting candidate set is passed to the parameter-refinement stage.

\subsection{Candidate Refinement}
\label{sec:orion_candidate_refinement}

The candidates surviving the initial reduction are passed to a more precise search using \texttt{PRESTO}. The candidate DM from \texttt{PULSCAN} is used to generate a dedispersed time series with \texttt{prepsubband}, while the estimated spin frequency guides the subsequent search with \texttt{accelsearch}. The DM is refined in the subsequent stages of the pipeline, allowing the final candidate parameters to be determined more accurately than in the initial search.

The initial \texttt{accelsearch} search is centred on the frequency estimates from \texttt{PULSCAN}, with the frequency window typically spanning $\pm10\%$ of the estimated frequency. If this search fails to produce an \texttt{ACCEL} candidate file, the frequency window is expanded to $\pm30\%$ to allow a broader search around the initial estimate. This provides a fallback when the narrower search does not yield a candidate file. For targeted searches, the acceleration or jerk search can likewise be restricted according to the expected parameter range.

Although \texttt{PULSCAN} is primarily designed for the detection of accelerated pulsars, it can also detect pulsars exhibiting low jerk values \citep{White_2025}. Leveraging this capability, \texttt{ORION} provides a user-defined flag to enable a frequency-constrained jerk search, allowing candidates from targeted searches to undergo an additional refinement stage when the expected binary parameters warrant a higher-order acceleration model. This option is particularly useful for compact or strongly accelerated binary systems. The analysis presented in this paper, however, uses the default configuration, with this flag set to ``OFF''.

\iffalse
For targeted searches, candidates can instead be passed through a constraint-based jerk search when the expected binary parameters warrant a higher-order acceleration model. This provides an additional refinement stage for compact or strongly accelerated binary systems.
\fi

\begin{figure*}
    \centering
    \includegraphics[width=0.48\textwidth]{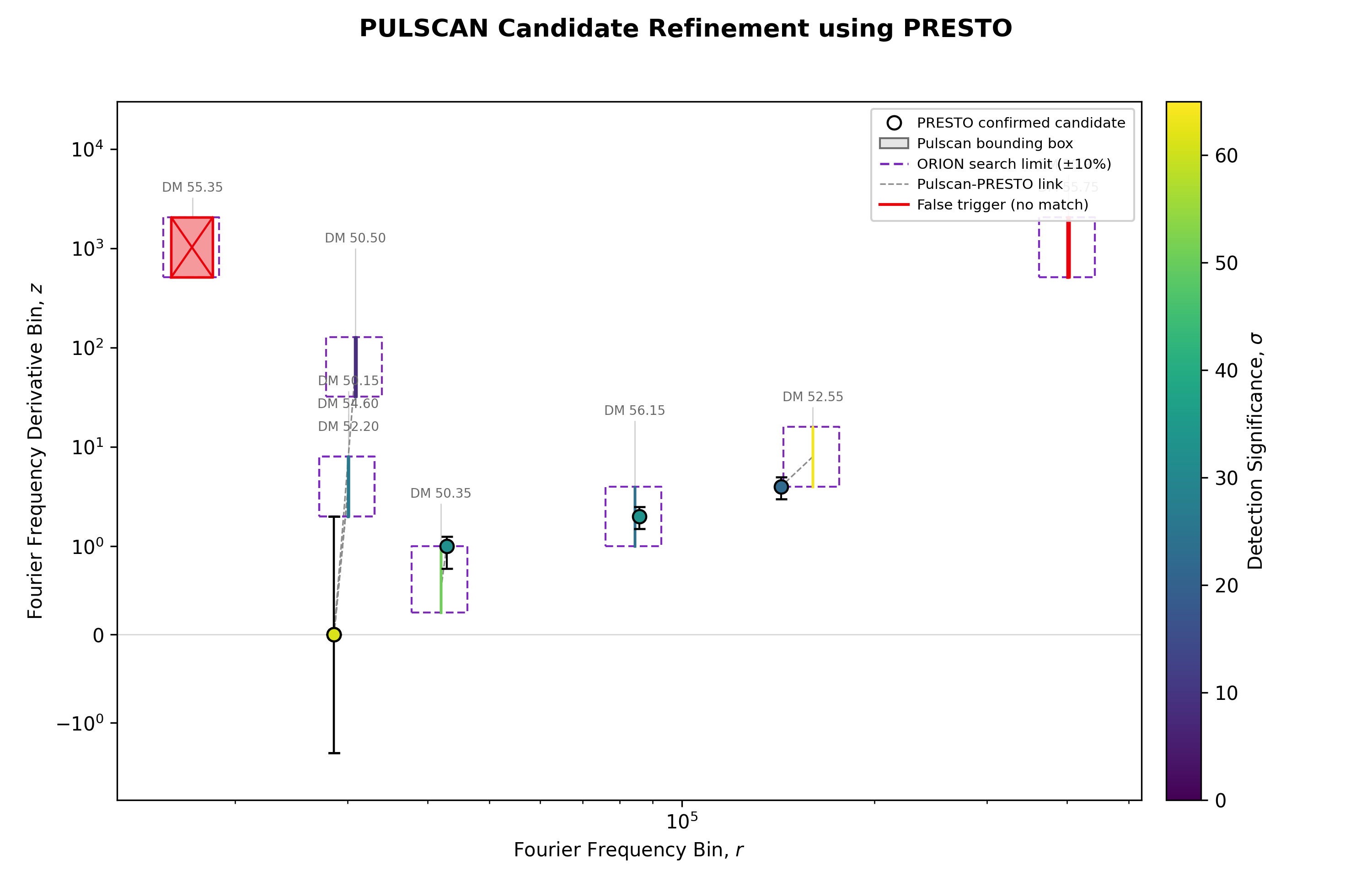}
    \hfill
    \includegraphics[width=0.48\textwidth]{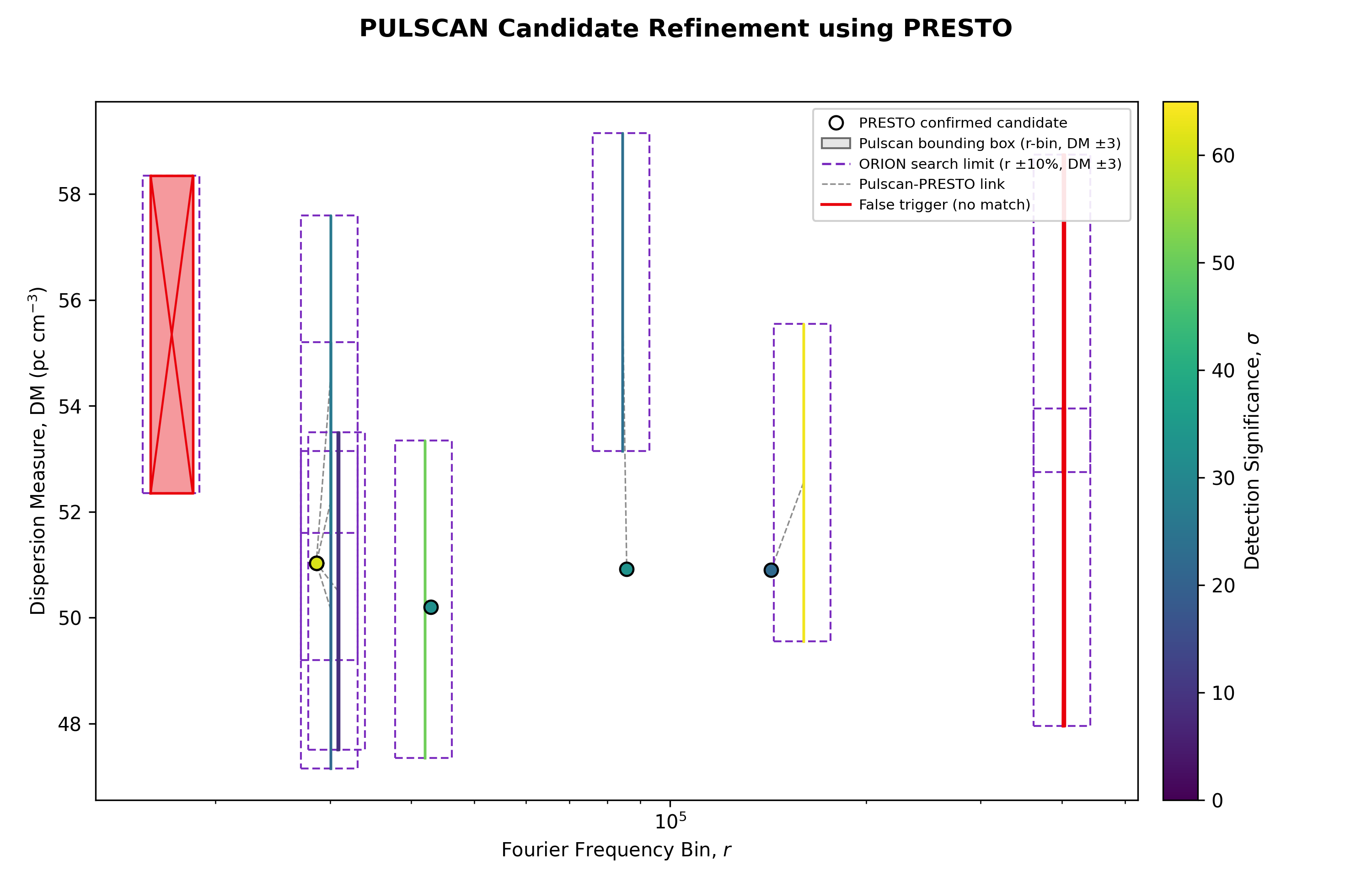}
    \caption{Refinement of \texttt{PULSCAN} candidates using localized \texttt{PRESTO} searches. The left panel shows the Fourier derivative-bin ($z$) versus Fourier-frequency-bin ($r$) parameter space, while the right panel shows the corresponding DM--$r$ distribution. The broad regions associated with the \texttt{PULSCAN} candidates define the initial search space, within which the more precise \texttt{PRESTO} solutions are identified.}
    \label{fig:PULSCAN_PRESTO}
\end{figure*}

Figure~\ref{fig:PULSCAN_PRESTO} illustrates the refinement in the frequency-domain parameter space. The broader regions returned by \texttt{PULSCAN} define the initial search uncertainty, while the \texttt{PRESTO} detections provide more precise frequency and acceleration-related parameters within these regions. The DM--frequency representation further shows the consistency of the refined solutions with the DM range identified by the initial search. By restricting the search to these localized regions, the refinement stage reduces the computational cost relative to repeating a full acceleration search over the original parameter space.

\subsection{Candidate Folding and Classification}
\label{sec:orion_folding}

The refined candidates are subsequently folded to generate diagnostic pulse profiles. \texttt{ORION} supports folding using either \texttt{PRESTO}'s \texttt{prepfold} or \texttt{PulsarX} \citep{Men_2023}. The resulting diagnostic products are passed to automated candidate classifiers, including PICS through \texttt{PulsarX}, before final manual inspection \citep{Men_2023}.

The overall design of \texttt{ORION} therefore  reserves the computationally expensive matched-filtering, jerk search, and folding operations for candidates that have already passed the high-throughput GPU search and multistage candidate filtering. This enables the processing of the large number of simultaneous beams while retaining the parameter precision required for subsequent candidate validation.

\section{Pipeline Validation and Performance}
\label{sec:orion_validation}

In this section, we validate the performance of the pipeline. We compare the single beam timing of \texttt{ORION} with \texttt{PSS}. \texttt{ORION} is tested on simulated binary pulsars and an on-sky validation is done with known GC pulsars.

\begin{figure}
\centering
\includegraphics[width=0.48\textwidth]{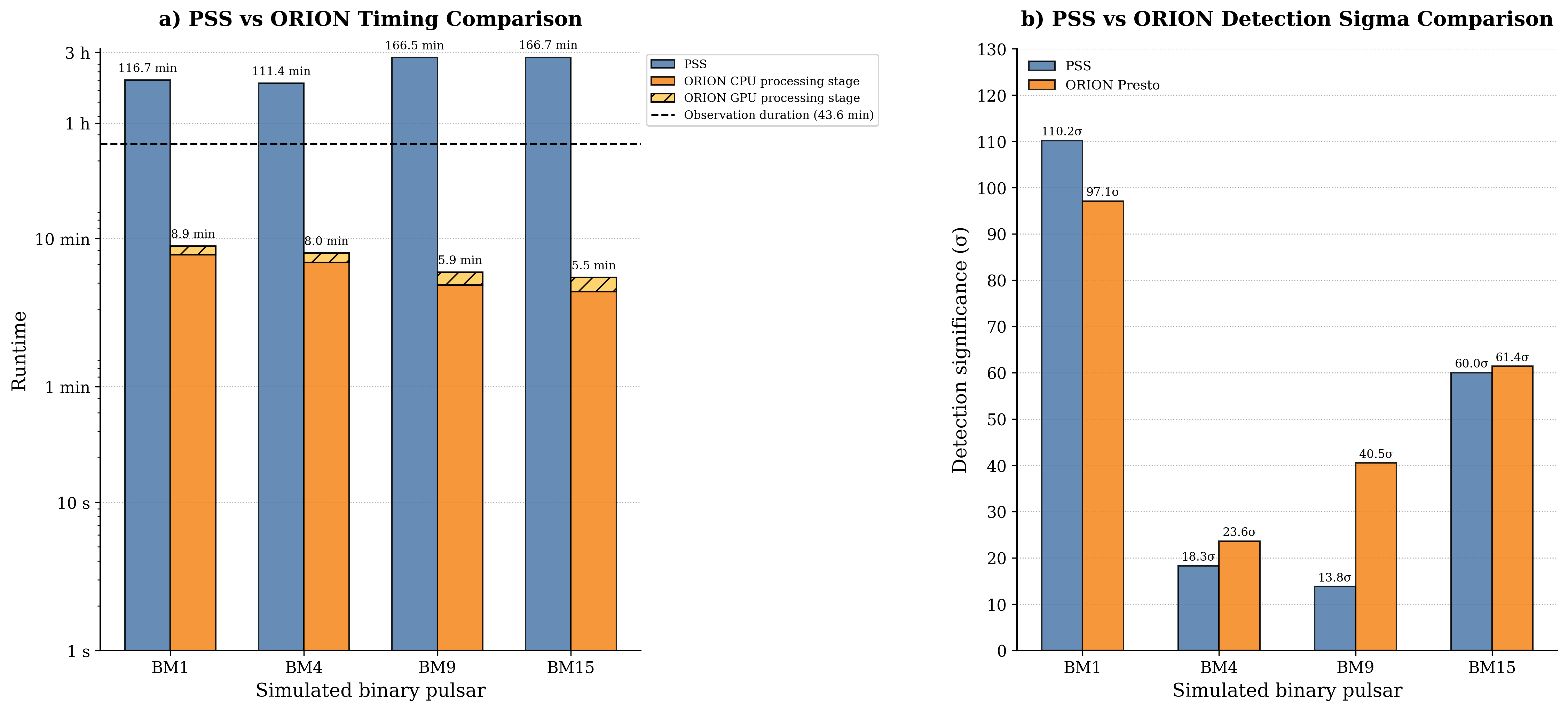}
\caption{Comparison of the processing time and detection significance of the Pulsar Search Script (\texttt{PSS}) and \texttt{ORION} for four simulated binary pulsars. (a) Processing time for each pipeline, with the observation duration and average \texttt{PSS} runtime indicated for reference. (b) Detection significance of the recovered pulsars. \texttt{ORION} substantially reduces processing time while achieving comparable detection significance}
\label{fig:ORION_vs_PSS}
\end{figure}

\iffalse
\begin{figure}
\centering
\includegraphics[width=0.48\textwidth]{Plots/ORION_vs_PSS_timing_sigma.pdf}
\caption{Comparison of the performance of the Pulsar Search Script (\texttt{PSS}) and \texttt{ORION} for four simulated binary pulsars. \texttt{ORION} provides a substantial reduction in processing time while yielding higher detection significance for all four simulated sources.}
\label{fig:ORION_vs_PSS}
\end{figure}
\fi

\begin{figure}
\centering
\includegraphics[width=0.48\textwidth]{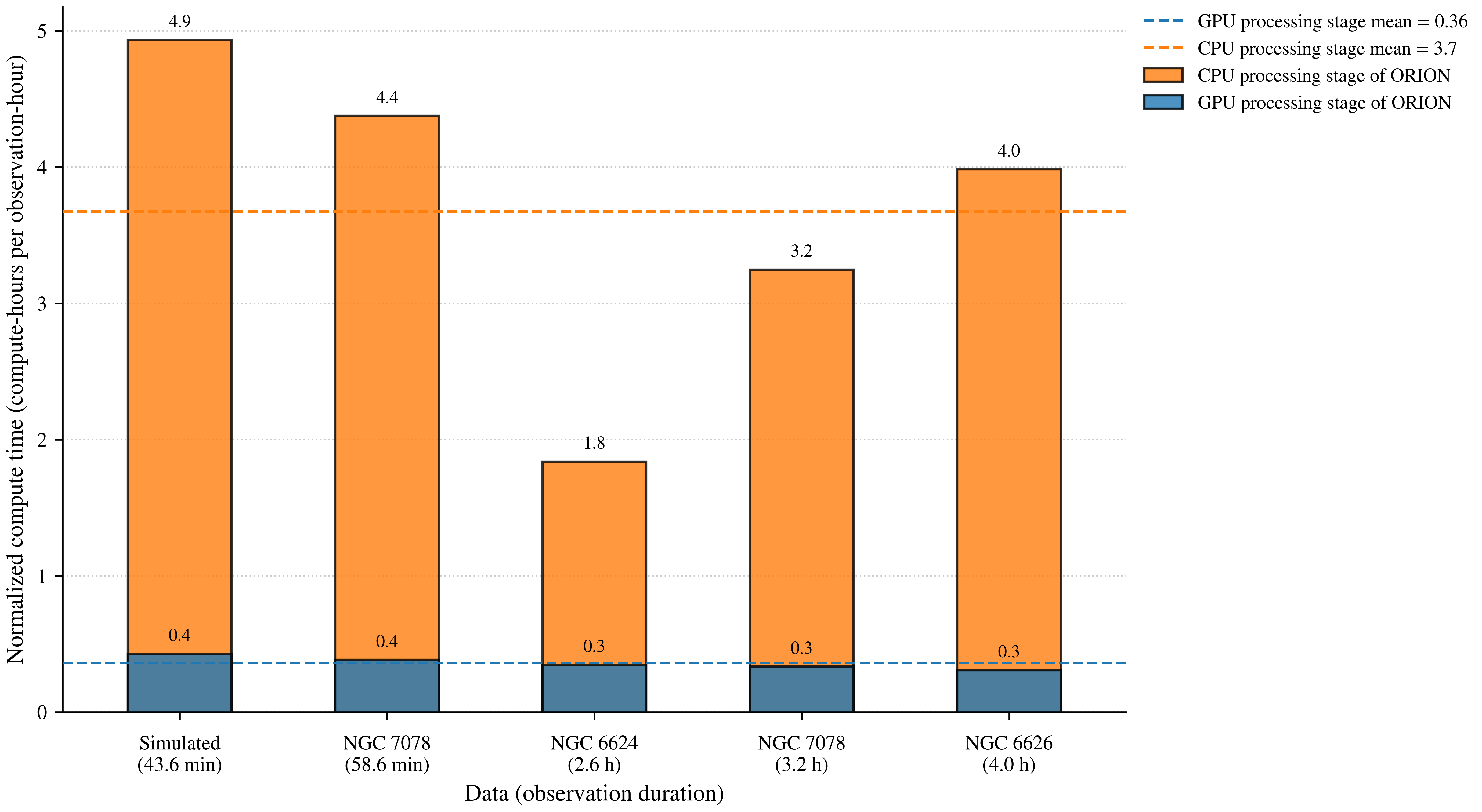}
\caption{Normalized computational cost of the \texttt{ORION} pipeline for different datasets, measured using the \texttt{ORION} computing cluster comprising 64 CPU cores and 8 NVIDIA A100 GPUs. The GPU processing stage includes \texttt{PULSCAN}, while the CPU processing stage comprises the frequency-localized, \texttt{PRESTO}-based acceleration search and \texttt{PulsarX}-based candidate folding. Compute times are normalized by the corresponding observation durations.}
\label{fig:ORION_compute_time}
\end{figure}

\subsection{Comparison with the GCGPS Phase I Pipeline}
\label{sec:orion_pss_comparison}

The performance of \texttt{ORION} is compared with the Pulsar Search Script (\texttt{PSS}) used for GCGPS Phase~I \citep{Das_2025, Das_2026}. \texttt{PSS} is a CPU-based pipeline built around \texttt{PRESTO} acceleration searches and was used for the Phase~I single-beam observations that led to the discovery of seven MSPs in four GCs \citep{Das_2025, Das_2026}. It therefore provides a useful reference for evaluating the computational performance of \texttt{ORION}. The computational cost of the two pipelines was measured using equivalent single-beam data. For \texttt{PSS}, the processing time includes the \texttt{PRESTO}-based dedispersion, acceleration search, candidate sifting, and folding stages. For \texttt{ORION}, the corresponding processing comprises the GPU-accelerated \texttt{PULSCAN} search, candidate reduction, \texttt{PRESTO}-based parameter refinement, and folding.

Figure~\ref{fig:ORION_vs_PSS} compares the processing time and detection significance obtained for four simulated binary pulsars. \texttt{ORION} provides a substantial reduction in processing time, achieving a speed-up of approximately 15--30 times over \texttt{PSS} in these tests. The detection significance of the recovered candidates is comparable between the two pipelines, with \texttt{ORION} yielding higher significance for three of the four simulated sources. These differences do not imply an intrinsic sensitivity advantage in the acceleration-search algorithm, as both pipelines use \texttt{PRESTO}'s \texttt{accelsearch} for candidate refinement; rather, they reflect differences in candidate selection.

A substantial fraction of the remaining \texttt{ORION} processing time is spent on the localized acceleration searches performed by \texttt{PRESTO} and candidate folding with \texttt{PulsarX}. The computational cost of these stages depends primarily on the number of candidates retained after \texttt{PULSCAN} and the subsequent \texttt{ORION} sifting stage. This number is also influenced by the dispersion-measure (DM) range and step size adopted in the search. All observations considered here are targeted searches conducted over a DM range of $\pm 5~\mathrm{pc\,cm^{-3}}$ around the target DM, with a DM step size of $0.1~\mathrm{pc\,cm^{-3}}$. Consequently, the number of candidates requiring further processing does not scale directly with observation duration, and no clear monotonic trend in the total processing time with duration is expected. In contrast, \texttt{PSS} processes a comparatively consistent number of candidates across these observations, resulting in less variation in its processing time. This highlights the importance of effective candidate reduction in further improving the computational efficiency of \texttt{ORION}.

To quantify the computational cost across different observation durations, we measured the normalized compute time of the principal \texttt{ORION} search stages. Figure~\ref{fig:ORION_compute_time} shows the compute-hours required per observation-hour for the GPU-accelerated \texttt{PULSCAN} and CPU-based \texttt{PRESTO} stages. The mean normalized costs are 0.36 GPU compute-hours and 3.7 CPU compute-hours per observation-hour, respectively. These values were measured using the \texttt{ORION} computing cluster, comprising eight GPUs and 64 CPU cores, and represent the resource requirements of the respective stages on this configuration. They should not be interpreted directly as relative wall-clock runtimes, since the stages use different processing hardware. The variation between observations reflects, in part, differences in the number of candidates passed to the subsequent processing stages rather than observation duration alone.

All performance tests were conducted using this cluster configuration of eight GPUs and 64 CPU cores. For the full 160-beam Phase~II data, the current implementation processes observations in approximately 0.25--0.33 times the observation duration in elapsed processing time with this configuration, corresponding to roughly three to four times real-time throughput. This demonstrates the potential of \texttt{ORION} to handle the increased computational demands of the multibeam survey.

\subsection{On-sky validation: recovery of known GC pulsars}
\label{sec:onsky_validation}

\begin{figure*}
    \centering
    \includegraphics[width=0.7\textwidth]{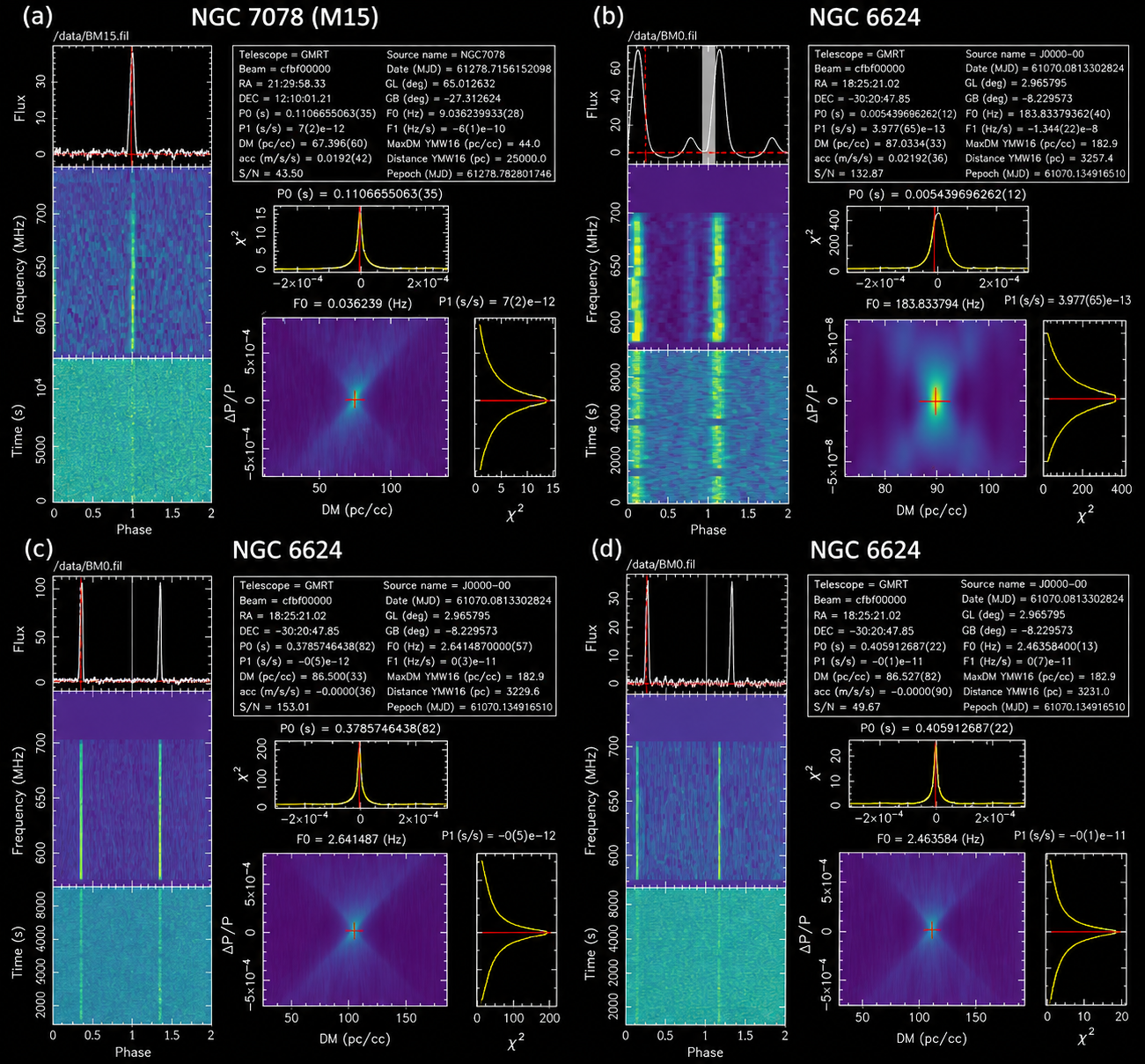}
    \caption{Pulsar detections obtained from Cycle 49 observations of the GCGPS Phase II survey. Panel (a) shows the detection of a pulsar in the globular cluster NGC 7078 (M15), while panels (b)--(d) show three pulsars detected towards NGC 6624. The corresponding spin periods are approximately 110.66 ms, 5.43 ms, 378.57 ms, and 405.91 ms, respectively.}
    \label{fig:ver0_output_plots}
\end{figure*}

The Phase~II search pipeline was validated on sky by processing Cycle~49 observations containing previously known globular-cluster pulsars. All known pulsars expected to be detectable in the validation observations were successfully recovered, comprising one pulsar in NGC~7078 (M15) and three pulsars toward NGC~6624, as shown in Figure~\ref{fig:ver0_output_plots}. This provides an end-to-end test of the search chain, including the beamformed data products, dedispersion, periodicity search, multibeam filtering, and candidate diagnostic generation.

The recovered pulsars span a wide range of spin periods, from approximately 5.4~ms to 406~ms, and are detected at their expected dispersion measures. The corresponding diagnostic plots show clear periodicity signatures in the folded profiles and phase--time and phase--DM diagrams, demonstrating that the pipeline successfully retains the characteristic signatures of both millisecond and longer-period pulsars. The recovery of sources with substantially different spin periods also provides a useful check that the search is not restricted to a narrow region of the pulsar parameter space.

These detections provide an important on-sky validation of the Phase~II processing pipeline before it is applied to the full GCGPS target sample. In particular, the successful recovery of known sources demonstrates that the beamformed observations can be processed through the complete search pipeline without losing established pulsar signals, while providing confidence in the candidate products used for subsequent inspection and confirmation.

\section{Scientific Implications}
\label{sec:scientific_implications}

As discussed in Section~\ref{sec:gc_dynamics}, MSPs in GCs are not confined to the cluster core. Pulsars at larger radii, out to and beyond the half-light radius, can include less-disturbed binaries and spider systems that received recoil kicks from the encounters that formed them, so a search restricted to the central region can miss a significant part of the population. The main contribution of the Phase~II strategy and \texttt{ORION} is to make spatially extended searches of GCs practical. Up to 160 synthesised beams cover a large fraction of a cluster in a single observation, and the GPU-accelerated, multibeam-aware processing keeps the computational cost manageable, at approximately 0.25--0.33 times the observation duration for the full 160-beam data.

The direct outcome of this approach is the discovery and recovery of pulsars, including binaries with approximately constant line-of-sight acceleration, across the full spatial extent of a cluster. Compared with the two phased-array beams of Phase~I, the dense beam layout increases the probability that a pulsar lies close to the centre of at least one beam, and the relative detection significances in neighbouring beams provide approximate positional information prior to interferometric localisation. The multibeam coincidence filtering also helps to separate localised celestial sources from interference that appears in unrelated beams. Together, these features improve the prospects of finding pulsars at larger cluster-centric radii, in clusters of large angular extent such as $\omega$~Centauri, and in clusters that have not previously been searched with comparable sensitivity. Since the current search assumes constant acceleration, highly jerked systems may still be missed; the planned Jerk \textsc{pulscan} extension (Section~\ref{sec:future}) is intended to address this limitation.

The wider scientific goals described in Section~\ref{sec:science_motivation} depend on follow-up observations that lie beyond the scope of \texttt{ORION}. Confirming candidates, determining orbital parameters and eccentricities, and measuring relativistic orbital effects and component masses all require long-term timing, supported by interferometric localisation. These measurements are what enable tests of gravity and constraints on neutron-star properties \citep{Freire_2011, Ridolfi_2021}, while timing-derived accelerations of a spatially complete pulsar sample can constrain the mass distribution of the cluster. By increasing the number of GC pulsars found across the whole cluster, including systems at large radii that are the most likely to be missed, \texttt{ORION} supplies the sample on which these follow-up studies depend.

\section{Future Developments}
\label{sec:future}

% \subsection{Jerk-aware Search with \texttt{Jerk\_PULSCAN}}
\label{sec:future_jerk}

The current implementation of \texttt{PULSCAN} uses boxcar filtering to identify signals undergoing approximately constant frequency drift. While computationally efficient, this approach limits its ability to characterize rapidly varying frequency drift and may reduce sensitivity to highly accelerated and jerked binary pulsars. Improving the searchable parameter space is important for discovering compact and exotic binary systems, particularly millisecond pulsars in dense globular-cluster environments.

To address this limitation, we plan to develop \texttt{Jerk\_PULSCAN}, which will extend the existing approach using a recursive boxcar filtering scheme to characterize variations in frequency drift and estimate the second derivative of the spin frequency, $\ddot{f}$. This extension is expected to improve the initial detection stage's sensitivity to highly jerked signals while retaining the computational efficiency of \texttt{PULSCAN}.

The estimates of spin frequency $f$, its first derivative $\dot{f}$, and its second derivative $\ddot{f}$ obtained from \texttt{Jerk\_PULSCAN} will then be used to constrain the corresponding parameter space in a subsequent Fourier-domain jerk search with \texttt{PRESTO}. Rather than searching the full $(f,\dot{f},\ddot{f})$ space blindly, the search will be restricted to a localized region informed by the initial candidate parameters. In particular, constraining the coupled frequency-derivative dimensions, including the relevant $z$- and $w$-bin ranges, is expected to reduce the number of trials required by the matched-filtering search.

This development will extend the coarse-to-fine strategy already implemented in \texttt{ORION} to jerk searches, combining efficient initial candidate detection with a more computationally demanding, localized refinement stage. By reducing the parameter-space volume explored during the jerk search, the approach is expected to enable computationally efficient searches for highly accelerated and jerked binary pulsars across the large number of synthesized beams in GCGPS Phase~II and future multibeam surveys, improving the prospects of discovering compact and dynamically exotic systems.

\section{Summary}
\label{sec:summary}

The Globular Cluster GMRT Pulsar Survey (GCGPS) aims to expand the known population of pulsars in Galactic globular clusters, which provide important laboratories for studying binary evolution, stellar dynamics, and fundamental physics. We have presented the motivation and observing strategy for GCGPS Phase~II, together with \texttt{ORION}, a GPU-accelerated multibeam pulsar search pipeline developed to process the resulting SPOTLIGHT observations.

GCGPS Phase~II extends the two-beam observing strategy of Phase~I by exploiting the capability of SPOTLIGHT to form up to 160 synthesised beams simultaneously. This substantially increases the instantaneous spatial coverage and allows the beam layout to be tailored to the angular extent of each cluster. In particular, the overlap between adjacent beams and the number of antennas contributing to each beam can be adjusted to balance spatial coverage against sensitivity. We also quantified the effects of intra-channel dispersion smearing and finite sampling for the proposed Phase~II configuration, motivating the adopted dispersion-measure limits of $\mathrm{DM}\leq 50~\mathrm{pc\,cm^{-3}}$ in Band~3 and $\mathrm{DM}\leq 150~\mathrm{pc\,cm^{-3}}$ in Band~4.

To address the computational demands of searching a large number of simultaneous beams, we developed \texttt{ORION}, which combines the high-throughput, unmatched Fourier-domain filtering of \texttt{PULSCAN} \citep{White_2025} with localized parameter refinement using \texttt{PRESTO}. Its staged workflow applies threshold, harmonic, and multibeam coincidence filtering before refining promising candidates and generating folded diagnostic products for classification and inspection. This approach reduces the need for computationally expensive searches across the full parameter space for every beam.

Tests using simulated binary pulsars show that \texttt{ORION} reduces processing time by approximately a factor of 15--30 relative to the CPU-based Pulsar Search Script (\texttt{PSS}) used in Phase~I, while achieving comparable or higher folded detection significance for the tested sources. These significance differences do not establish an intrinsic sensitivity advantage, as both pipelines use \texttt{PRESTO}'s \texttt{accelsearch} for candidate refinement. On-sky validation using Cycle~49 observations recovered four known globular-cluster pulsars, with spin periods spanning approximately 5.4--406~ms. All known pulsars expected to be detectable in these validation observations were recovered, providing an end-to-end test of the search and candidate-selection stages.

With 8 GPUs and 64 CPU cores, the current implementation processes full 160-beam Phase~II observations in approximately 0.25--0.33 times the observation duration, corresponding to roughly three to four times real-time throughput. These measurements demonstrate the feasibility of processing the increased data volume associated with the Phase~II multibeam strategy using the current computational configuration.

Together, the Phase~II observing strategy and \texttt{ORION} provide a framework for expanding pulsar searches across the spatial extent of globular clusters while keeping the computational cost manageable. Future development of \texttt{Jerk\_PULSCAN} is intended to improve the initial characterization of signals with rapidly varying frequency drift and constrain subsequent jerk searches.

\section*{Acknowledgments}
We acknowledge the invaluable contribution of colleagues from the Oxford e-Research Centre (OeRC), NVIDIA, and the Centre for the Development of Advanced Computing (C-DAC). We thank our colleagues at the C-DAC for their support in setting up the Param Brahmand data centre at the GMRT. We acknowledge the funding of the SPOTLIGHT backend (called Param Brahmand) under the National Supercomputing Mission's (NSM) Phase 3, as well as the support of the Department of Atomic Energy (DAE), Government of India, under project no. 12-R\&D-TFR5.02-0700, for the contributions towards the overhead cost. The uGMRT is operated by the National Centre for Radio Astrophysics of the Tata Institute of Fundamental Research, India.

We gratefully acknowledge support from the ”Building Indo–UK Collaborations Towards the Square Kilometre Array”, funded under the DAE–STFC Technology and Skills Programme, which facilitated the development of SPOTLIGHT's highly efficient, real-time transient search pipelines. We sincerely thank the uGMRT engineers involved in the SPOTLIGHT survey for their relentless efforts in commissioning and stabilising SPOTLIGHT's correlator and beamformer systems, ensuring high quality data flow for all of SPOTLIGHT's real-time and offline systems. We also thank the uGMRT operators for their coordinated efforts in successfully conducting the SPOTLIGHT survey observations. This research was supported in part by the International Centre for Theoretical Sciences (ICTS) for the FTSky: A program in the field of Fast Radio Transients (code: ICTS/FTSky2025/10). ND gratefully acknowledges the financial support provided by Premji Invest’s generous donation to TIFR to pursue the application of AI to accelerate scientific discovery in astrophysics. This funding helped support a Research Assistant position at NCRA–TIFR and enabled this research.

\bibliography{bibliography}
\bibliographystyle{pasa}

\end{document}